\documentclass[aps,floatfix,a4paper,prb,showpacs]{revtex4}
\usepackage{graphicx}
\usepackage{amssymb,amsmath}

\begin{document}

\title{\bf Reply to 
Comment on ``Electron transport through correlated molecules  computed using the 
time-independent Wigner function: Two critical tests''}

\author{Ioan B\^aldea}
\email{ioan.baldea@pci.uni-heidelberg.de}
\altaffiliation[Also at ]{National Institute for Lasers, Plasma, 
and Radiation Physics, ISS, POB MG-23, RO 077125 Bucharest, Romania.}
\author{Horst K\"oppel} 
\affiliation{Theoretische Chemie,
Physikalisch-Chemisches Institut, Universit\"{a}t Heidelberg, Im
Neuenheimer Feld 229, D-69120 Heidelberg, Germany}

\begin{abstract}
In their Comment, Greer et al
(i) put us in charge of a pretended wrong claim, which we never made in 
Phys.~Rev.~B {\bf 78}, 115315 (2008), where we criticized a method (DG)
proposed by two of them, 
(ii) incorrectly claim that the DG method 
can reproduce the conductance quantum $g_0$, but (iii) to deduce $g_0$
for a toy model, they carry out calculations within the standard Landauer method, 
which has nothing to do with the DG's.
We present results for their model obtained within the DG method,
which demonstrate that the DG method 
fails as lamentably as in the examples we presented in our earlier work. 
We also analyze the physical reasons why the DG method fails.
\end{abstract}


\maketitle

\renewcommand{\topfraction}{1}
\renewcommand{\bottomfraction}{1}
\renewcommand{\textfraction}{0}
\section{Introduction}
\label{sec-introduction}
In their Comment,\cite{GreerComment:09} 
Greer, Delaney, and Fagas (henceforth GDF) claim that they can resolve 
the conundrum expressed in our paper \cite{Baldea:2008b} (referred to as I below),
where we challenged a method (DG) put forward by two of them in Ref.~\onlinecite{DelaneyGreer:04a}
and subsequently used in their Refs.~\onlinecite{DelaneyGreer:04b,DelaneyGreer:06,Fagas:07}.
Unfortunately, their attempt to refute our criticism of I (a reference which GDF 
even omit in their bibliography list) is based on a series of inaccuracies and  
statements without any support. This starts from the abstract of Ref.~\onlinecite{GreerComment:09},
wherein GDF incorrectly state that 
we criticized a so-called MECS (many-electron correlated scattering)
approach, which was not even 
mentioned in I, nor defined in the works
\onlinecite{DelaneyGreer:04a,DelaneyGreer:04b,DelaneyGreer:06,Fagas:07} 
that made the object of our criticism. 
Before discussing the controversial issues in detail, we emphasize the following:

(i) Using the acronym MECS, GDF obviously 
distract attention from the fact that the (DG) method
\cite{DelaneyGreer:04a,DelaneyGreer:04b,DelaneyGreer:06,Fagas:07}
envisaged by us is manifestly a \emph{variational} one. 
In their ``brief introduction 
to the MECS approach'' (Sect.~II of Ref.~\onlinecite{GreerComment:09}), they completely omit 
to state that the \emph{variational} ansatz (namely, the minimization of the total 
energy in the presence of an external bias $V$) is an essential ingredient of the DG method;
in that introduction, they do not define a \emph{transport} approach, 
but merely review well known properties of the 
Wigner function (WF) $f(q, p)$. 

(ii) With this acronym and ``carefully'' chosen quotations from I, by
only mentioning in passing 
``\emph{the variational structure of the MECS calculations performed to date \ldots}'' (Sect.~V),
without noting that \emph{all} their calculations criticized in I are \emph{variational},
they aim to convey the false impression that the DG method merely consists of open
boundary conditions (OBCs) expressed in terms of the WF, and that I only criticizes these 
WF-OBCs. 
We stress from the very beginning that what we criticized
in I is the DG \emph{variational} approach as such,
and in particular the WF-OBCs in the \emph{specific} context of that 
method employed to a finite cluster; the DG approach, as any transport theory, 
cannot be reduced to whatever boundary conditions (BCs).

(iii) None of the results of GDF (their Sect.~IV) 
were obtained within the DG method, but simply within the well-established 
Landauer approach for an uncorrelated toy model (GDF model hereafter).\cite{toy}

GDF mainly claim that: our critique in I 
of the DG method is invalid, and they present calculations within a simple model 
that reproduce the conductance quantum $g_0$.
The first claim is wrong, as explained in Sect.~\ref{sec-wrong-claim}. 
In Sect.~\ref{sec-not-dg-method}, we show that the second claim is irrelevant, because 
their derivation of $g_0$ is done within the  
Landauer approach
and not within the variational DG approach criticized by us.
Some challenges to the results thus obtained by GDF are presented in Sect.~\ref{sec-challenges}.
Next, in Sect.~\ref{sec-dg-to-gdf} we present results 
for the GDF model deduced within the DG method, which clearly demonstrate
the failure of that method. The other issues raised by GDF are addressed in Sect.~\ref{sec-further}. 
In Sect.~\ref{sec-fail} we indicate the 
physical reasons why the DG method fails. Conclusions are presented in Sect.~\ref{sec-conclusion}.
\section{Refuting the GDF criticism to our work}
\label{sec-wrong-claim}
The main presumptions made by GDF to refute the criticism of I 
are that:
(i) we incorrectly presumed an asymmetric injection of momentum, 
(ii) we incorrectly assumed that there is no chemical potential difference when applying 
OBCs, and 
(iii) we appear to be confusing in the application of OBCs when using either 
momentum or energy distributions.

All these claims are wrong.
To arrive at our basic Eqs.~(19-21), which yielded the results presented in Figs.~2---5, and 7,
demonstrating the lamentable failure of the DG method, we made \emph{exactly} 
what that method prescribes:
energy minimization, Eq.~(1), with WF-OBCs, Eqs.~(5, 6), wave function normalization, Eq.~(9), and, 
if not otherwise specified, current conservation, Eq.~(8). More specifically:

To (i): A \emph{momentum} asymmetry was assumed 
neither in the analytical results of Eqs.~(19-21), nor in the numerical results of Figs.~2---5, and 7. 
In our WF-OBCs, Eqs.~(5, 6), the ``momenta''
$p_{L,R}$ in the LHS and RHS are the \emph{same}. Otherwise, we should have written, e.g.,
$f(q_{L,R}, p_{L,R} \pm \delta p) = f_0(q_{L,R}, p_{L,R})$, $\delta p = \mathcal{O}(V)$. 

To (ii): This is wrong, we do assume a nonvanishing chemical potential difference ($eV$). 
The bias $V$ enters our calculations, because the quantity to be minimized of Eq.~(1) does depend on it via $W$. 
Consequently, the wave function $\Psi$ obtained within the DG procedure does contain an asymmetry 
between the left and right electrodes for $V \neq 0$, and this asymmetry reflects itself 
in the WF at the boundaries. 
Without this dependence, e.~g., the quantity $\delta f = f - f_{0} \neq 0$ 
for electrons flowing from the device into electrodes displayed in Fig.~2  
would have been zero.

To (iii): By faithfully applying the DG method to well-defined models in I, 
we were simply faced with a \emph{mathematical} problem whose solution is \emph{exact} 
and \emph{unique} within the 
linear response theory, and unambiguously yields unphysical results. 
We needed not explicitly discuss, 
e.~g., the distinction between momentum and energy distributions.

The numerical results presented in I, which are completely unphysical, are nothing 
but the results of the variational DG method, because  in I we did \emph{nothing} else than 
exactly what that method prescribes.
Therefore, any critique to I represents in reality a critique of the DG method itself.
\section{Rebuttal of the claim on the reproduction of the conductance quantum}
\label{sec-not-dg-method}
GDF write: ``\emph{In practice, the}'' WF ``\emph{is used to constrain the momenta flow out of the electron reservoirs and into the scattering region}'' (Sect.~II). 
This ``practice'' does not obviously refer to the Comment; simply, GDF use Eqs.~(3--5) nowhere
in calculations (not even in their envisaged works).
Relevant for the present debate would have been to apply the variational DG approach
(including its original WF-OBCs) and see whether it is able to correctly describe the 
transport,
and not to show that another (Landauer's) approach 
can reproduce correct results (conductance or WF).

In the Comment, GDF do nothing else than apply the standard 
Landauer approach to deduce the conductance [see their Eqs.~(12) and (13)]
and to compute the WF in electrodes' middle.
So, they cannot pretend that the \emph{variational DG} approach 
correctly reproduces the conductance of their 
simple uncorrelated model. Their statement in Introduction that in ``\emph{Sect.~IV 
a calculation of conductance quantization \ldots is given using the MECS construction}'' 
is not true. Equally wrong is their claim
that the BCs ``\emph{as formulated in MECS applied to}'' their ``\emph{model reproduces the 
well-known result of conductance quantization}''.
We did not challenge the Landauer approach, and the fact that
using it GDF can deduce a correct result ($g_0=e^2/h$) 
for an uncorrelated model, which is trivial from the point of view of the Landauer approach, 
is not at all astonishing:
the transmission through a barrier of vanishing height ($V\to 0$) is equal to unity. 
This correct result, deduced within a correct approach, has absolutely no relevance for the 
validity of the variational DG method, and our critique of the latter is not in the 
least affected. In the GDF calculations, BCs are imposed not by using the WF but by simply 
postulating standard scattering forms of the asymptotic 
single-electron wave functions [Eqs.~(7) and (11)]. They use these forms, in a kind of 
a posteriori check, to show that the WFs computed within 
the Landauer approach, in and out of equilibrium, 
at carefully chosen boundaries behaves as expected physically.
Computing the WF within the Landauer approach (and \emph{not} within the DG's) and 
claiming that ``\emph{indeed it is observed that in the single particle case
constraining the incoming momentum inflow via the Wigner function implies solving
the Schr\"odinger equation \ldots}'' (Sect.~IV.A of GDF) is totally unfounded. 
The GDF calculation represents in itself a logical inconsistency 
hard to understand: within Landauer calculations, they assume 
Fermi distributions (shifted in energy by $\mu_R - \mu_L = eV$ for $V\neq 0$)
in electrodes, and show that 
the WFs (approximately) just look like the Fermi functions already postulated.
Within this philosophy, one can solely infer that the Landauer approach is indeed 
self consistent, but this is trivial and irrelevant for the validity of the DG's.

As noted in Introduction, GDF wish to reduce the whole debate to the WF-OBCs, 
but no transport theory can be solely reduced to a certain type of BCs. 
In I we criticized the WF-OBCs
in the \emph{specific} context of the variational DG method, and 
did not challenge the WF-OBCs in general.
See the end of the third paragraph on page 2 of I and the lines after Eqs. (3) and (4) in I], or 
what we explicitly wrote (last paragraph of I): 
\emph{``While the idea of formulating boundary conditions in terms of the Wigner function for 
correlated many-body systems is interesting, the manner in which it was imposed in''} 
Ref.~\onlinecite{DelaneyGreer:04a} \emph{``turns out to be inappropriate.''}

By deducing a simple result within the Landauer approach and not within the DG's,
GDF raise very serious doubts that they can also derive the correct result within 
the latter. Confirming these doubts, we shall demonstrate in Sect.~\ref{sec-dg-to-gdf} 
that for the simple model of their choice, the DG approach fails as lamentably as in 
the cases presented in I.
\section{Challenges to the GDF calculations}
\label{sec-challenges}
The GDF's Landauer calculations affect in no way our critique to the DG method.
However, especially because GDF attempt to present these results as if 
they were related to Ref.~\onlinecite{DelaneyGreer:04a},
we note several errors of Ref.~\onlinecite{GreerComment:09}.

For the GDF model, the transmission can be 
trivially obtained analytically, $T(\varepsilon) = 4 k_L k_R /(k_L + k_R)^2$, 
$k_L \equiv \sqrt{2 m \varepsilon}/\hbar$, and 
$k_R \equiv \sqrt{2 m(\varepsilon - e V)}/\hbar$,
\cite{Constantinescu} and it yields the following exact expression of the current 
\begin{eqnarray}
\displaystyle
I(V) & = & 
-\frac{e}{h} \frac{8\varepsilon_F}{n_F^2}\sum_{n_L \geq n_V}^{n_F}\frac{n_L^2 x_R}{\left(n_L + x_R\right)^2}
+ \frac{e}{h} \frac{8\varepsilon_F}{n_F^2}\sum_{n_R=1}^{n_F}\frac{x_L n_R^2}{\left(x_L + n_R\right)^2}
\stackrel{L\to \infty}{\longrightarrow}
V g_{L}(eV/\varepsilon_F), \nonumber \\
g_{L}(x) & \equiv & g_0 \frac{8}{3x^3}\left[
\frac{x^3}{2} + \sqrt{1 + x} - \sqrt{1 - x} + x
\left( \sqrt{1 + x} + \sqrt{1 - x} - 3 \right)
\right];  \lim_{x\to 0} g_{L}(x) = g_0, \label{eq-I} \\
x_{L,R} & \equiv & \sqrt{n_{R,L}^2 \pm n_{V}^2}, n_V\equiv n_F \sqrt{e V/ \varepsilon_F} . \nonumber
\end{eqnarray}
The above $n_V$ coincides with the RHS of GDF's Eq.~(16). 

Instead of giving these simple formulas, GDF present some qualitative considerations
aiming to show that (i) the conductance quantum is correctly reproduced, and 
(ii) this result has something 
to do with the results and the electrode sizes ($L\approx 2$\,nm) of their controversial 
works.\cite{DelaneyGreer:04a,DelaneyGreer:04b,DelaneyGreer:06,Fagas:07}
To deduce $g_0$ in Sect.~IV.B, they approximate $T(\varepsilon) \simeq T(\varepsilon_F)
\simeq 1$; this amounts to \emph{implicitly} 
assume the linear response limit ($e V \ll \varepsilon_F$).
In the last paragraph of Sect.~IV.A, GDF claim that their considerations to deduce $g_0$ 
``\emph{apply well to electrode lengths as small as $1$\,nm}''($=L/2$).
They assume that $n_F$ is large and $\Delta k = 2\pi/L$ is small
[this should mean that $n_F = k_F L/2\pi \gg 1$ and $\Delta k = 2\pi/L\ll k_F$, cf.~Eq.~(7) of GDF]. 
For $L=2$\,nm and their completely \emph{unphysical} 
value $k_F = 0.12$\,nm$^{-1}$ (cf.~Fig.~3 of Ref.~\onlinecite{GreerComment:09}), 
one gets $n_F = 0.0038$ (!) and $\Delta k = 3.14$\,nm$^{-1} \gg k_F$(!). 
In reality, to mimic gold electrodes 
(Fermi velocity $v_F \simeq 1.4 \times 10^6$\,m/s, $\varepsilon_F = m v_F^2/2 \simeq 5.6$\,eV), 
one needs a $k_F$-value \emph{hundred} times larger,
$k_F \simeq 12$\,nm$^{-1}$. Even then, one gets $n_F=3.8$, and satisfying the 
above conditions is problematic.
In fact, to derive Eq.~(18) from Eq.~(17) in Ref.~\onlinecite{GreerComment:09} GDF need not a large 
$n_F$, but rather a large $n_V$, which should obviously be much smaller than $n_F$.
If we admit that $n_V=2$ is a number ``much'' larger  than unity and 
``much'' smaller than the other ``large'' number $n_F\alt 4$, Eq.~(\ref{eq-I}) leads to 
$V \sim 1$\,volt. 

\begin{figure}[htb]
\centerline{\includegraphics[width=0.25\textwidth,angle=-90]{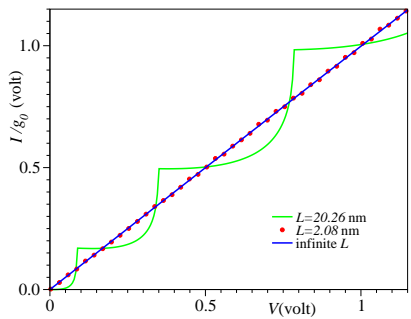}}
\caption{\label{fig:jToy} 
I-V characteristics obtained exactly from Eq.~(\ref{eq-I}) 
for infinite electrodes with $k_F=12$\,nm$^{-1}$ (as in gold), 
and for two finite sizes $L$ (given in the legend).}
\end{figure}

How poor is the linear approximation in this range and how bad is the description 
based of electrodes with $L=2$\,nm, exactly the linear sizes of the 
$Au_{13}$ and $Au_{20}$ clusters used in GDF's ab initio 
works,\cite{DelaneyGreer:04a,DelaneyGreer:04b,DelaneyGreer:06,Fagas:07}
can be seen in Fig.~\ref{fig:jToy}, and any further comments are superfluous.
If, as in the present case, the exact current for $L\to \infty$ is known, one may still 
argue that the ``correct'' trend can be unraveled by a ``clever'' inspection of the curve 
for very short electrodes ($L=2$\,nm) in Fig.~\ref{fig:jToy}. However, one may more 
legitimately ask how reliable could be considered 
a theoretical result obtained with very short electrodes 
in implementations for complex realistic cases (like those of 
Refs.~\onlinecite{DelaneyGreer:04a,DelaneyGreer:04b,DelaneyGreer:06,Fagas:07}) 
where neither the exact values nor the 
electron spectrum details are known a priori. 
From the curve for $L=2$\,nm of Fig.~\ref{fig:jToy} one deduces 
a linear conductance ($V \to 0$) of $\sim 10^{-2} g_0$, i.~e., hundred times smaller than the 
true value $g_0$. 

\begin{figure}[htb]
\centerline{\includegraphics[width=0.25\textwidth,angle=-90]{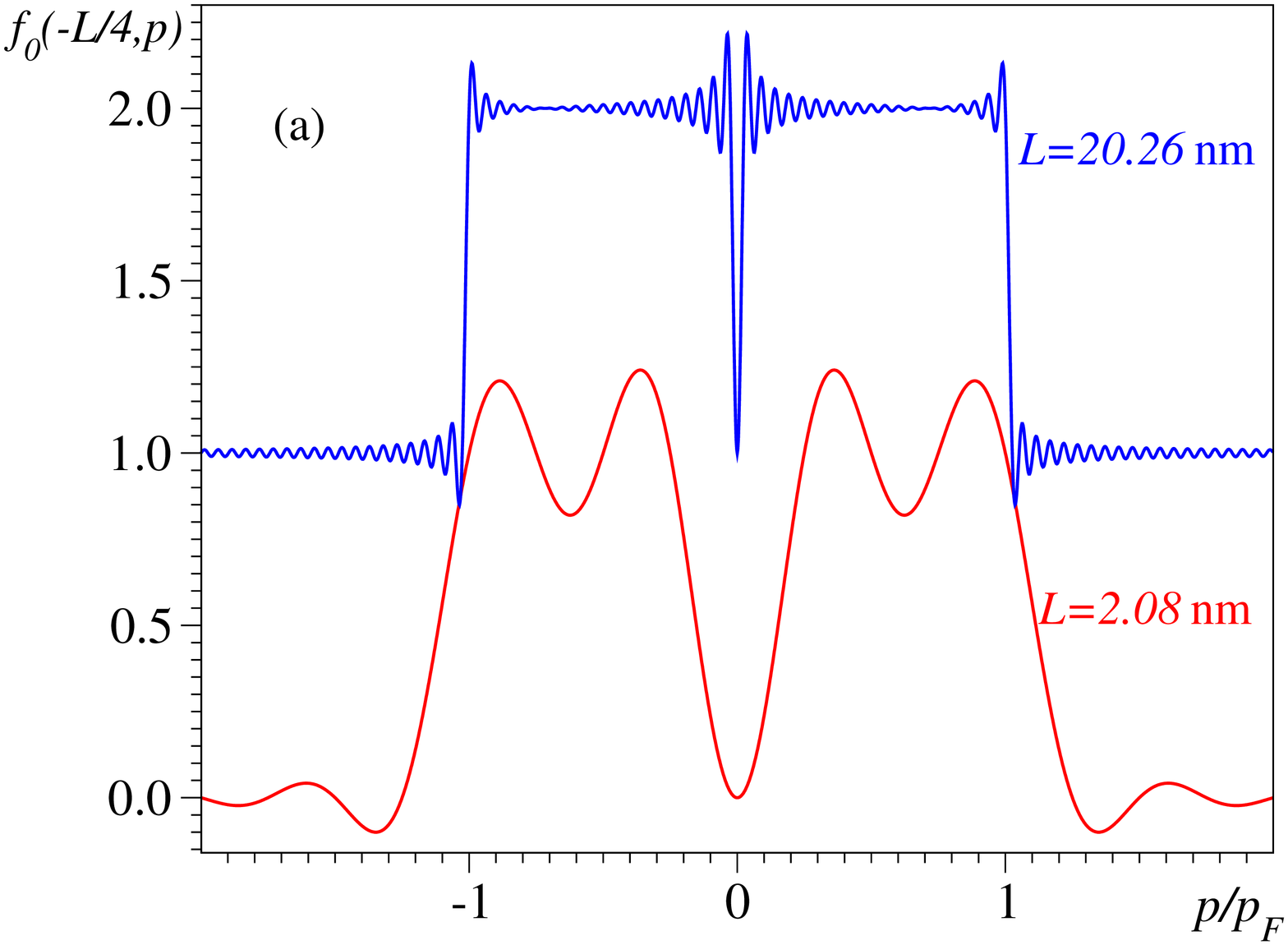}
\hspace*{2ex}{\includegraphics[width=0.25\textwidth,angle=-90]{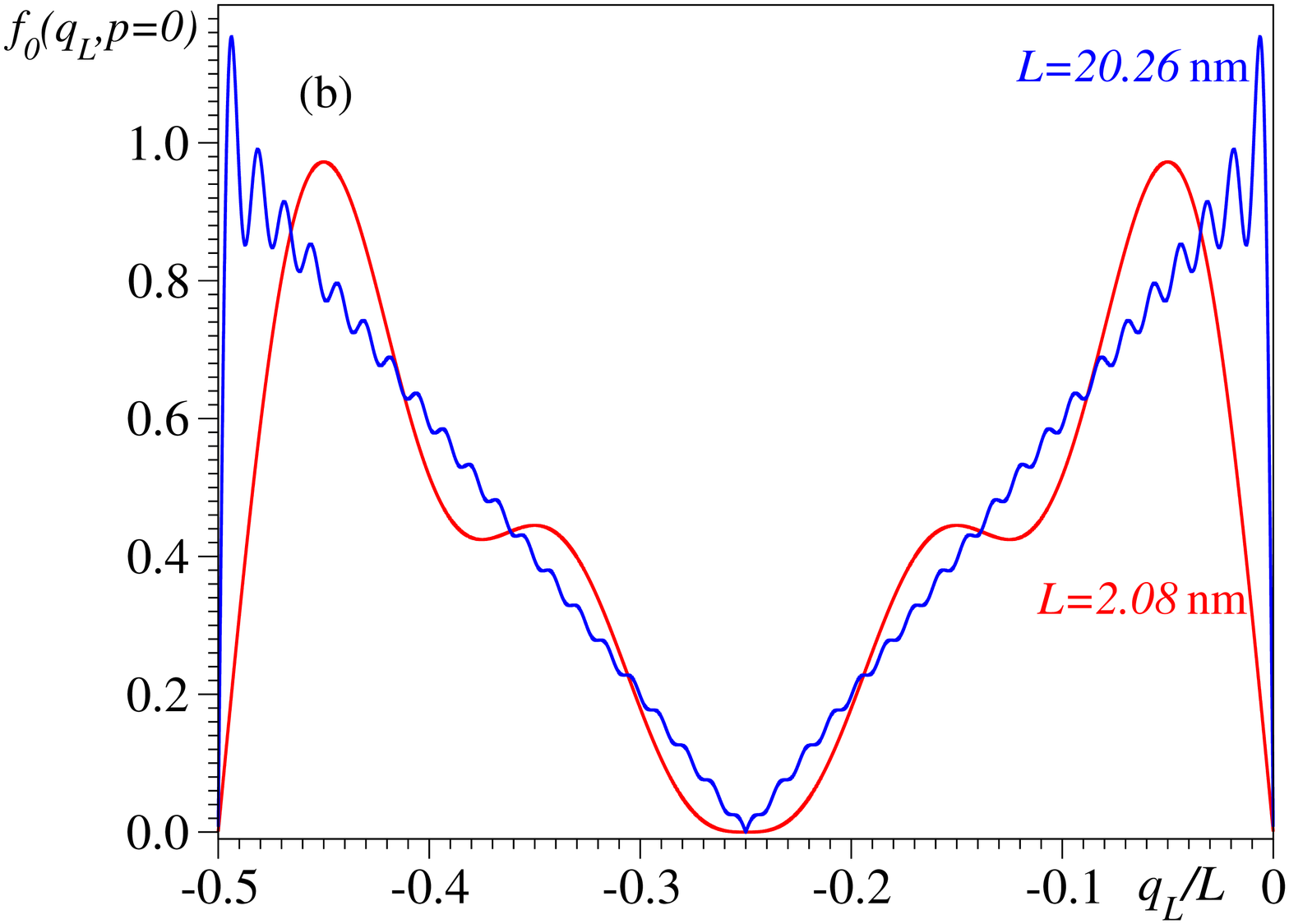}}
\hspace*{2ex}{\includegraphics[width=0.25\textwidth,angle=-90]{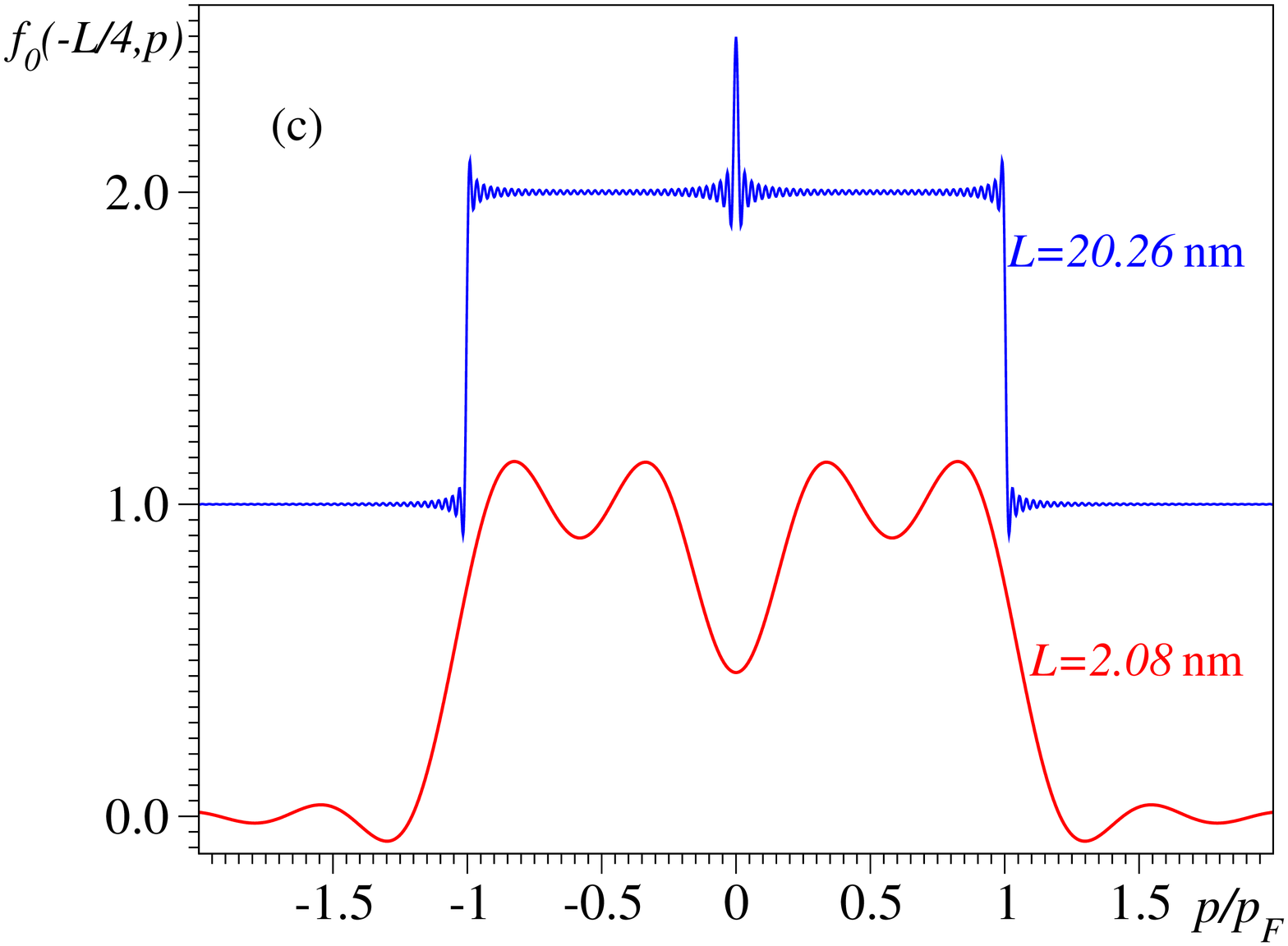}}}
\caption{\label{fig:wf-gdf-bk} 
Equilibrium Wigner function computed: 
(a, b) using the GDF formulas, and (c) for a confined system  
for two sizes $L$ (given in the legend) and $k_F = p_F/\hbar = 12$\,nm$^{-1}$.}
\end{figure}

As their results on the WFs obtained within the Landauer approach  
are irrelevant for the DG validity, we restrict ourselves here to a few critical remarks.
In Sect.~II, GDF mention that the WF, 
as a function of {energy} defined in the phase space, tends 
rapidly toward the Fermi distribution with increasing number $N$ of particles in a 
{confining} potential. Indeed, in their Ref.~9, Cancellari et al showed that 
$\mathcal{F}(\varepsilon_p) \sim \int_{\varepsilon_p, \varepsilon_p + \delta \varepsilon_p} 
d\,\mathbf{q}_{1}^{N}
 d\,\mathbf{p}_{1}^{N}
f^{(N)}(\mathbf{q}_1^{N}, \mathbf{p}_1^{N})$ 
behaves as the Fermi distribution $f(\varepsilon_p)$. Notice that the system of that Ref.~9 
is \emph{confined} (while GDF use \emph{periodic} BCs, $k_n = 2 \pi n/L$), and 
Cancellari \emph{et al} attribute a physical meaning to a function whose argument is 
the \emph{energy} $\varepsilon_p$, which is unequivocally defined in the uncorrelated case discussed by GDF,
and \emph{not} to a function of \emph{momentum} $p$. Contrary to them, GDF ascribe a physical meaning 
to $f(q_{L,R}, p)$, i.~e., at fixed locations (without $q$-integration) 
and interpret $p$ as a physical momentum 
of electrons, e.~g.,~which move toward right for $p>0$ and toward left for $p<0$.
We do not analyze here $f(q, p)$ at $V\neq 0$.\cite{Baldea:unpublished}
We only point out an error in the Comment for $V=0$. 
The exact expression of their Eq.~(10) deduced from their formulas for finite $L$ is
\begin{eqnarray}
\displaystyle
f_{0}(-\frac{L}{4}, p) & = & \frac{2}{L}\sum_{n=1}^{n_F}
\left\{
\frac{\sin\left[\left(p/\hbar + k_n\right)L/2\right]}{p/\hbar + k_n} +
\frac{\sin\left[\left(p/\hbar - k_n\right)L/2\right]}{p/\hbar - k_n}
\right\} \nonumber \\
 & = & \frac{\sin \pi P}{\pi}\sum_{n=1}^{n_F}
(-1)^{n}\left(\frac{1}{P + n} + \frac{1}{P - n}\right) ,
\label{eq-WF-0}
\end{eqnarray}
where $P \equiv n_F p /(\hbar k_F)$. Inspecting Eq.~(\ref{eq-WF-0}), one can immediately see
that the RHS vanishes for $p=0$. Therefore, their formula and the 
curves for $L=2$\,nm and $L=20$\,nm of their Figs.~3a and b are incompatible 
(the case $q_R=+L/4$ is similar).
The Wigner function computed by means of Eq.~(\ref{eq-WF-0}), i.~e., GDF's \emph{own} formula 
for their sizes is shown 
in Fig.~\ref{fig:wf-gdf-bk}a.\cite{round-L} As visible there, the dip around $p=0$ [of a width 
$\delta p = 2 \hbar k_F/n_F \propto 1/L$, cf.~Eq.~(\ref{eq-WF-0})], 
becomes narrow only at larger $L$. If GDF used their formula to compute the WF at $q_{L,R}=\mp L/4$, 
they  would have shown the curve of our Fig.~\ref{fig:wf-gdf-bk}a, not too much resembling a Fermi 
distribution for $L=2$\,nm, which 
mimics their gold electrodes.\cite{DelaneyGreer:04a}

Limitations of the physical meaning of the WF are well known.\cite{mahan,frensley:90,Datta:97}
Noteworthy, in emphasizing these limitations even at equilibrium, 
textbooks (see, e.~g., Ref.~\onlinecite{mahan}, ch.~3.7, pp.~202-203) 
choose typical examples \emph{just} like the one 
invoked by GDF (cf.~Sect.~IV.A) to claim that the WF is physically useful.
The fact that the boundary locations whereat the WF was
computed by GDF cannot be arbitrary and should be ``carefully'' chosen
(just in electrodes' middle to GDF, $q_{L,R}=\mp L/4$) already raises warning bells  
on its appropriateness for applying BCs. Generalizing Eq.~(\ref{eq-WF-0}) away from $q_{L,R} = \mp L/4$ 
by restricting the integration $r$-range to either electrode in the GDF formulas is straightforward.
To illustrate the sensitivity of $f_{0}(q_L, p)$ to the boundary location, 
we depict in Fig.~\ref{fig:wf-gdf-bk}b the component $p=0$, a classical textbook's example
(e.~g, ch.~8.8, p.~327 of Ref.~\onlinecite{Datta:97}), which defies a simple physical interpretation.
Noteworthy, our results of the \emph{genuine} DG calculations in I are quite 
unphysical even with this choice (boundaries \emph{exactly} in electrodes' middle):
this is just the choice in Fig.~7, this choice also qualitatively changes nothing in Fig.~3.  
\section{Results of the variational DG method for the GDF model}
\label{sec-dg-to-gdf}
We proceed by showing what GDF should have shown, namely the results of the variational 
DG method for the chosen model. To obtain the linear response of the GDF model, 
the working equations (5-21) of I (which yield an \emph{exact} and 
\emph{unique} solution) 
deduced from the faithful implementation of the variational 
DG method without any other extra assumption (cf.~Sect.~\ref{sec-wrong-claim}), 
can be straightforwardly used. Only a minor adaptation is needed in Eqs.~(7), (10), and (12), namely to 
consider a continuous space coordinate $-L/2 < x < L/2$
instead of a discrete lattice, and this will be briefly indicated below. 
An important strong point of the 
approach of I is that it allows to determine self-consistently the limit of its applicability,
namely the highest bias $V < V_{lr}$ compatible with the 
linear response approximation. For this, the expansion coefficients $A_n = \mathcal{O}(V)$ 
[$\Psi = A_0 \Psi_0 + \sum_{n\neq 0} A_n \Psi_n$, $A_0 = 1 - \mathcal{O}(V^2)$]
should satisfy the condition 
$\mathcal{S} \equiv \sum_n \vert A_n\vert^2 \ll 1$. 
As a practical numerical criterion 
we imposed $\mathcal{S}=0.01$, which typically yielded $V_{rl} \sim 1$\,volt. Therefore, 
all numerical results presented in this section for $V\neq 0$ 
are for $V=1$\,volt.

Without applied bias ($V=0$), an eigenstate $k$ of 
each electron out of the $N$ noninteracting electrons considered is characterized by a wave
function $\phi_{k}(x)$ and an energy $\varepsilon_{k}$ satisfying the Schr\"odinger equation
\begin{equation}
\label{eq-1p}
h(x)\phi_{k} (x) \equiv \left[-\frac{\hbar^2}{2m}\frac{d^2}{d\,x^2} + v(x)\right] 
\phi_{k}(x) = \varepsilon_{k} \phi_{k} (x) .
\end{equation}
The second quantized electric current $\hat{j}(x)$ and Fano $\hat{F}(q, p)$ operators 
needed to compute the linear response read
\begin{equation}
\displaystyle
\hat{j}(x) = \sum_{k,k^{\prime}} J_{k,k^{\prime}}(x) c_k^{\dagger} c_{k^{\prime}}; 
J_{k,k^{\prime}}(x) = i\frac{e\hbar}{2 m}
\left[
\phi_{k}^{\ast}(x) \frac{\partial\phi_{k^{\prime}}(x)}{\partial x} -
\frac{\partial\phi_{k}^{\ast}(x)}{\partial x} \phi_{k^{\prime}}(x) 
\right] ;
\end{equation}
\begin{equation}
\displaystyle
\hat{F}(q, p) = \sum_{k,k^{\prime}} F_{k,k^{\prime}} (q, p) c_k^{\dagger} c_{k^{\prime}}; 
F_{k,k^{\prime}} (q, p) = \int d\,r
\phi_{k}^{\ast}(q-r/2) \phi_{k^{\prime}}(q+r/2) e^{-i p r/\hbar} .
\end{equation}
The action of an applied voltage $V(x)$ is expressed by 
\begin{equation}
\displaystyle
\hat{W} = \sum_{k,k^{\prime}} V_{k,k^{\prime}} c_k^{\dagger} c_{k^{\prime}}; 
V_{k,k^{\prime}} = -e \int_{-L/2}^{L/2} d\,x
\phi_{k}^{\ast}(x) \phi_{k^{\prime}}(x) V(x) .
\end{equation}
For the GDF potential, the antisymmetric form 
$V(x) = V/2$ if $-L/2 < x < 0$, and $V(x) = -V/2$ if $0 < x <L/2$ is more convenient, 
because it permits to separate the even ($\Psi_{g}$) and odd ($\Psi_{u}$)
many-body eigenstates, as discussed in I. For this, it is necessary to further assume a 
symmetric potential barrier $v(-x)=v(x)$, which can be also included 
to make the calculations a bit more realistic. This (let us call it statical) barrier 
should not be confused with the barrier $V(x)$ related to the applied bias. For the GDF 
model, $v(x)\equiv 0$, and $\varepsilon_{k}=\hbar^2 k^2/(2m)$. 
Within the 
DG approach, the system is confined within $-L/2 < x < L/2$, 
$\phi_{k} (\pm L/2) \equiv 0$, and therefore 
$\phi_{k_n}^{g}(x) = \sqrt{2/L} \cos(k_n x)$ and 
$\phi_{k_n}^{u}(x) = \sqrt{2/L} \sin(k_n x)$, where $k_n\equiv\pi n/L$.
For even eigenstates (superscript $g$), $n=1,3,5,\ldots$ is odd, 
while for odd eigenstates (superscript $u$), $n=2,4,6,\ldots$ is even. 
In the ground state $\Psi_0$, the highest occupied single electron level has 
$k=k_F=\pi N/L$.
GDF's electron orbitals $\phi_{k_n}(x) = \exp(i k_n x)/\sqrt{L}$ 
satisfy periodic BCs, $k_n=2\pi n/L$, where $n$ is a signed integer 
($-n_F \leq n \leq n_F$). 
Without asking whether it makes sense to consider systems which are not 
confined within the original variational DG approach, for completeness 
and for comparison with the exact results for the GDF model (Sect.~\ref{sec-challenges}), 
we carried out calculations for both cases 
(termed confined and periodic below). 

Because the operators $\hat{W}$, $\hat{j}$, and $\hat{F}(q,p)$ 
are bilinear, \emph{all} their matrix elements 
$\langle \Psi_{n}\vert \ldots c_k^{\dagger} c_{k^{\prime}}\vert\Psi_0\rangle$ 
needed as input into the working Eqs.~(12)---(21) of I can be obtained exactly. 
All the exact excited many-electron eigenstates $\Psi_n$ that 
contribute are particle-hole excitations,
$\vert \Psi_{n}\rangle \to \vert \Psi_{k,k^{\prime}} \rangle = 
(-1)^{S_{k,k^{\prime}}} c_k^{\dagger} c_{k^{\prime}} \vert \Psi_0\rangle$. Here, the sign factor
accounts for the ordering adopted to fill the Fermi sea.  
So, exactly as in I, we can present exact full CI calculations done within the 
DG approach, and the results should be exact if that approach were valid. 

Within its variational scheme, the DG approach determines the steady state 
$\Psi$ at $V \neq 0$ by constraining the WF of incoming electrons at the electrode-device 
boundaries (at $q_{L,R} = \mp L/4$, as by GDF) to that of the ground state $\Psi_0$
\begin{equation}
f(q_{L,R}, \pm p>0)  \equiv 
\langle \Psi\vert \hat{F}(q_{L,R}, \pm p>0) \vert \Psi \rangle= 
\langle \Psi_0\vert \hat{F}(q_{L,R}, \pm p>0) \vert \Psi_0 \rangle 
\equiv f_0(q_{L,R}, \pm p>0) .
\end{equation}
The constrained minimization of the total energy allows the system ($ -L/2 < x < L/2$) 
to optimize the WF of outgoing electrons and therefore the differences below 
are allowed to be nonvanishing
\begin{equation}
\label{eq-delta-f}
\delta f(q_{L,R}, \mp p>0) \equiv
f(q_{L,R}, \mp p>0)  - f_0(q_L, \mp p>0) =
\langle \Psi\vert \hat{F}(q_{L,R}, \mp p>0) \vert \Psi \rangle -
\langle \Psi_0\vert \hat{F}(q_{L,R}, \mp p>0) \vert \Psi_0 \rangle 
\neq 0 .
\end{equation}
Due to this fact, by mathematical construction, the DG steady state 
breaks the time reversal.
In Fig.~\ref{fig:wf-gdf-bk}c, we present results for the WF 
$f_{0}(q_L, p) = f_{0}(q_R, p)$ in the ground state $\Psi_0$ of the confined system. 
Comparing panels $a$ and $c$ of Fig.~\ref{fig:wf-gdf-bk} one can see that, as expected 
for any 
physically relevant quantity, the equilibrium WF saturates at 
sufficiently large sizes $L$ and becomes independent on the BCs, let they be periodic or open. 
Notice that these results are for equilibrium ($V=0$) and have 
nothing to do with the DG method.

\begin{figure}[htb]
\centerline{\includegraphics[width=0.25\textwidth,angle=-90]{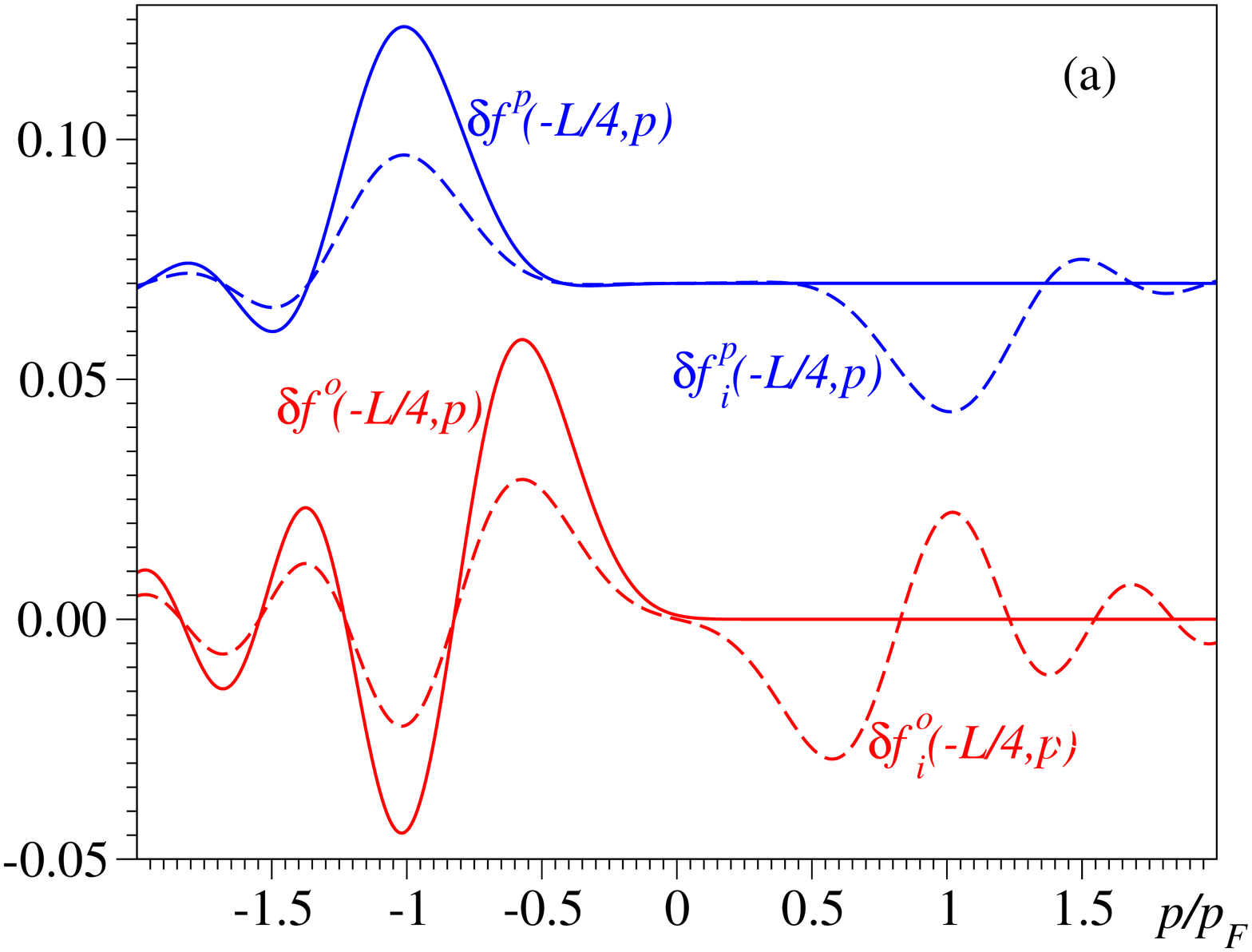}
\hspace*{2ex}\includegraphics[width=0.25\textwidth,angle=-90]{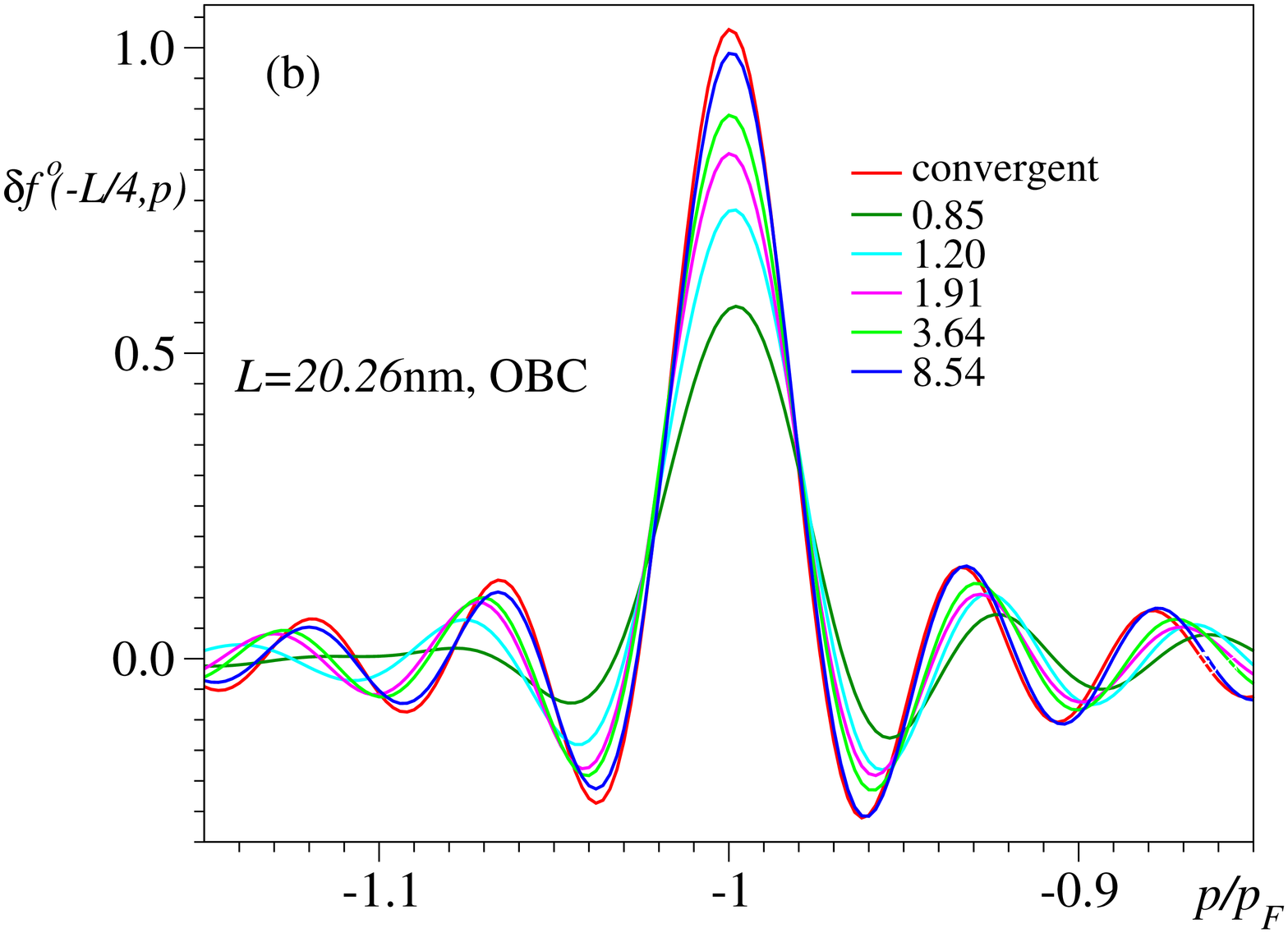}
\hspace*{2ex}\includegraphics[width=0.25\textwidth,angle=-90]{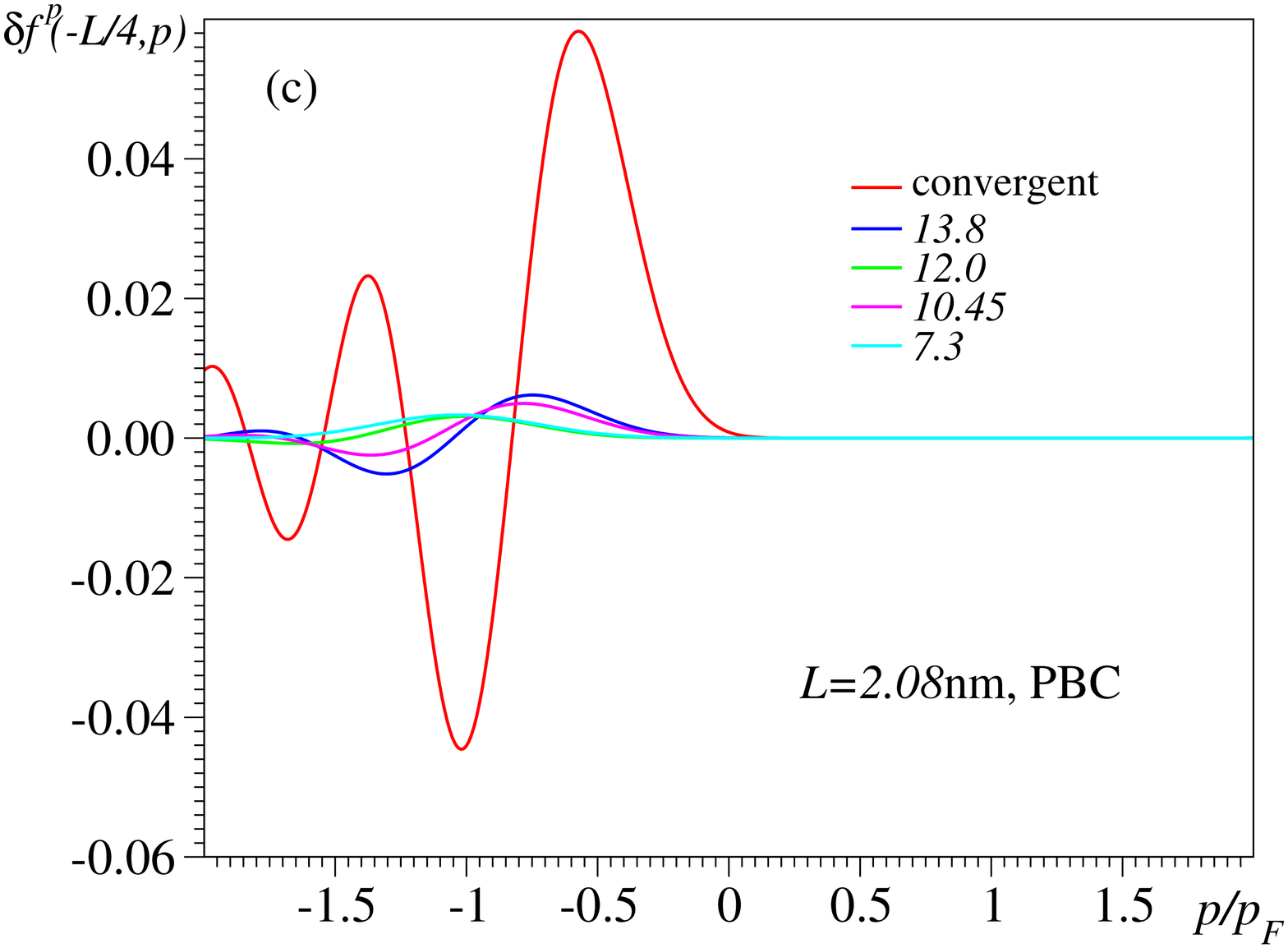}}
\caption{\label{fig:df-dfi-obc-pbc-bk} 
(a) Deviation from equilibrium 
of the Wigner function at the boundaries $\delta f(-L/4,p) = - \delta f(+L/4, -p)$ 
for the confined and periodic cases (superscripts $o$ and $p$, respectively) 
[see the main text and Eq.~(\ref{eq-delta-f})]; (b, c)
The very slow convergence of $\delta f(-L/4, p)$ 
by increasing all the exact multielectronic eigenstates 
up to high excitation energies $\Delta \varepsilon = r\varepsilon_F$ ($r$ is given in inset)
substantially larger than the metallic electrode 
bandwidth ($\sim \varepsilon_F$).
Because of rapid oscillations, only the range around the Fermi momentum $p_F$ 
is shown in (b).}
\end{figure}

Let us now examine the numerical results \cite{DG-numerical} 
for the differences of Eq.~(\ref{eq-delta-f}) computed within the DG approach
(Fig.~\ref{fig:df-dfi-obc-pbc-bk}).
Being for outgoing electrons that are unconstrained, they are indeed nonvanishing. 
Because $\delta f(q_{L}, p<0)$[$=\delta f(q_{R}=-q_{L}, -p)$ for linear response] 
are real, we note that both the real and imaginary parts of $A_u$'s of I contribute 
to $\delta f(q_{L,R}, \mp p>0)$. The latter contribution, denoted by $\delta f_{i}$,
does not vanish, as visible in Fig.~\ref{fig:df-dfi-obc-pbc-bk}a, which reveals that 
Im$A_u$'s related to the current [cf.~Eqs.~(18) and (21) of I] do not vanish. This holds both for 
the confined and for the periodic case.
So, in principle, the DG approach can allow a current flow.
The crucial point is now whether the electric current associated with the time reversal 
driven (or, ``mimicked'', cf.~Sect.~\ref{sec-fail}) by these $\delta f \neq 0$ 
is appropriately described within the DG approach. Our exact calculations 
demonstrate that, as in the two cases presented in I, the DG approach is invalid:
although $\delta f(q_{L,R}, \mp p>0)\neq 0$ and Im$A_u\neq 0$, 
both in the confined and the periodic cases, 
imposing current conservation or not,
the linear conductance $g_{DG}$ computed within the DG scheme 
vanishes within numerical accuracy or, to be on the safe side, $g_{DG} < 10^{-7} g_0$ 
for the investigated sizes $2$\,nm$\alt L \alt 400$\,nm.\cite{DG-numerical,x-p-grids}

In fact, this result is not at all astonishing; on the contrary, it should have been expected. 
The GDF model is nothing 
but the continuous version of the discrete model of Eq.~(22) of I,
for the particular choice $t_L = t_R = t_{d,L} = t_{d,R}$,
$\varepsilon_L = \varepsilon_R = \varepsilon_{g}$, $U=0$. As seen on the SWF-curve in 
Fig.~3 of I,  
at resonance ($\varepsilon_L = \varepsilon_R = \varepsilon_{g}$), the 
DG-conductance also vanishes. In this context, what is worth emphasizing is not 
that the DG approach yields a(n almost) vanishing conductance. In fact, the present calculations 
confirm the analysis Sect.~VI of I that the situation at resonance 
is particularly favorable for the DG approach to predict a vanishing current. 
But, as shown in I, $g_{DG}$ can also be nonvanishing. Fig.~3 of I depicted
situations where $g_{DG} \neq 0$.
But, completely unphysically, as seen in that figure, the farther away from resonance, 
the larger is the conductance 
$g_{DG}$. Although more tedious, exact DG calculations 
are also straightforward for the continuous-space counterpart of the uncorrelated discrete model
of I: to this, one should include 
in Eq.~(\ref{eq-1p}) a statical rectangular barrier 
$v(x) = \varepsilon_{g} \theta(\vert x\vert - l)$, wherein $\theta$ is the Heaviside function 
and $0 < l < L/4$.\cite{Baldea:unpublished} 
As far as the exact Landauer approach is concerned, the only difference from the uncorrelated 
model of I is that
the Lorentzian decay (cf.~Fig.~3 of I) of the exact transmission (Landauer conduction) 
away from resonance there is replaced by an exponential 
decays with $\varepsilon_g$ and $l$ here. 

To demonstrate the failure of the DG approach, we have chosen above the same method pursued in I,
of faithfully applying the DG method. Due to its extreme simplicity, the GDF model 
also offers another alternative decisive way to challenge the DG method.
Owing to the fact that the analytical exact scattering solutions of the 
single particle Schr\"odinger equation are known for \emph{arbitrary} $V$, 
one can pursue a different route.
Namely, one should consider small deviations of the single electron wave functions with 
respect to the exact ones ($\phi_k \to \phi_k + \delta \phi_k $), 
and investigate their impact on the DG-functional 
$\mathcal{H} \equiv 
\langle \Psi\vert \hat{H} + \hat{W}\vert \Psi\rangle - \omega \langle \Psi \vert\Psi\rangle - 
\sum_{p_L > 0} \lambda_{p_L} \langle \Psi\vert\hat{F}(q_L, p_L) \vert \Psi\rangle - 
\sum_{p_R < 0} \lambda_{p_R} \langle \Psi\vert\hat{F}(q_R, p_R) \vert \Psi\rangle
- \sum_j \xi_j \langle \Psi\vert\ (\hat{j}(x_j) - \hat{j}(0)) \vert \Psi\rangle$;
it should be minimum if the DG method were correct. 
That is, one should examine whether 
$\delta \mathcal{H} = \mathcal{O}(\delta \phi_k)$ 
or $\delta \mathcal{H} = + \mathcal{O}((\delta \phi_k)^2)$.
The expressions thus obtained 
are somewhat similar to those worked out in our approach of the DG linear response,\cite{Baldea:2008b}
but are obviously no longer limited to the linear response.\cite{Baldea:unpublished} 
Moreover, one can thus directly scrutinize the validity of 
the variational approach \emph{itself}, by constraining  
instead of the WF other properties, which are less sensitive to the boundary locations 
and, highly desirably, have a clearer physical meaning.
Thus, one can indeed exploit another ``\emph{advantage of an analysis based upon an analytical model \ldots}'' 
that ``\emph{avoids issues associated with linear response approximations, perturbation 
theory, variational methods in a finite basis, specific implementations of electronic structure, 
or other numerical approximations,
thereby allowing a clear focus on the physical assumptions made when using the}'' DG ``\emph{method}''
(cf.~Sect.~V of GDF).

The results obtained as described above deserve a separate analysis, which is beyond 
the scope of this Reply. 
What is important for the present purpose is the unambiguous demonstration that 
the variational DG method (without \emph{any} other assumptions) 
lamentably fails even for the model chosen by GDF themselves.
\section{Further issues}
\label{sec-further}
In Sect.~V, GDF claim that we criticized ``\emph{the use
of a configuration expansion to describe transport properties.}''
Such a statement cannot be found in I. 
GDF should not reduce the transport theories based on a configuration expansion (CI)
to the DG method. 
What they actually mean is our clear statement that, 
\emph{even} if the DG method were sound, 
it would be impractical. We showed that 
even if very many exact eigenstates were included, the current within the DG approach 
would very slowly converge; for a correlated system,
the first 300 exact eigenstates out of a total of 1225 are insufficient. 
One may think that the convergence is an issue \emph{only} for correlated systems. 
Indeed, in correlated systems, for which the DG method was designed, the convergence is extremely poor.

But let us examine the convergence in the uncorrelated GDF model. Because the current 
(practically) vanishes, let us inspect the quantity $\delta f(q_{L,R}, \mp p>0)$ that is 
directly related to it via time-reversal breaking. 
To exemplify, along with the convergent 
results, we present in 
Figs.~\ref{fig:df-dfi-obc-pbc-bk}b and c those obtained by including all \emph{exact} eigenstates 
with excitation energies $\Delta \varepsilon$ below a given value 
$\Delta\varepsilon \leq r \varepsilon_F$.
Both for sizes where ab initio DG computations were attempted ($L=2$\,nm),\cite{DelaneyGreer:04a}
and for those ($L\simeq 20$\,nm) where they are hopeless, \emph{all exact eigen}states 
with very high excitation energies ($r \gg 1$)
much larger even than the metallic electrode bandwidth ($\sim \varepsilon_F$) 
have to be included to reach convergence. Definitely, this were a 
serious challenge even for trivial uncorrelated systems and even if this approach were valid.
From the perspective of the severe limitations of the number $\mathcal{N}$ 
of multielectronic configurations 
in nontrivial ab initio calculations, the very slow convergence 
of the DG-results observed for the GDF model, for the two models of I, and many others 
\cite{Baldea:unpublished} also becomes an important issue 
because, e.~g., increasing 
$\mathcal{N}$ by $\sim 50$\,\% at the limit of feasibility 
and obtaining a change $\sim 3$\,\% in the investigated properties, one can erroneously 
conclude that the results almost converged. 
Indirectly referring to the poor convergence of the DG method demonstrated in I, 
GDF claimed in Sect.~V that ``\emph{integrated quantities such as the energy may be better 
approximated compared to local properties such as \ldots current density}''. 
But in reliable transport treatments the convergence is needed just for the latter.
Letting alone 
that, if current conservation were correctly accounted for, the (position-independent) 
current would also be an ``integrated quantity'',
$I = L^{-1}\int d\,x j(x)$, let us illustrate the effect of a truncated CI for a case 
relevant just for the electrode size 
$L/2\simeq 1$\,nm of Ref.~\onlinecite{DelaneyGreer:04a}. 
To compare the convergence of the DG method to that of a standard calculation, 
we consider the change in the 
total energy caused by $V\neq 0$ in the DG-state, 
$\Delta E^{DG}\equiv \langle \Psi\vert \hat{W}\vert \Psi\rangle$,
and in the state $\Psi^{(1)}$ obtained 
in the first order [$\mathcal{O}(V)$] of the perturbation theory 
(without any WF- and current-constraints) 
$\Delta E^{(1)}\equiv \langle \Psi^{(1)}\vert \hat{W}\vert \Psi^{(1)}\rangle$.
By including \emph{all exact eigen}states up to an excitation energy 
$\Delta \varepsilon = 8$\,eV,
we got within obvious notations $\Delta E^{DG}_{approx}= 0.18 \Delta E^{DG}_{conv}$ and 
$\Delta E^{(1)}_{approx}= 0.94 \Delta E^{(1)}_{conv}$. The convergence is an issue for the DG method,
not for a certain particular model. Notice that 
(i) we considered all aforementioned exact eigenstates, (ii) 
this $\Delta\varepsilon$-value is four 
times larger than that of the states considered relevant by DG,\cite{DelaneyGreer:04a} 
for which DG mentioned (without any detail) an inaccuracy factor $\sim 3$,\cite{DelaneyGreer:04a}
and (iii) the convergence dramatically deteriorates for a real correlated system. 
Contrary to DG,\cite{DelaneyGreer:04a} we cannot see any justified manner to evaluate the
inaccuracy factor related to a CI truncation from Figs.~\ref{fig:df-dfi-obc-pbc-bk}b and c. 

It is obvious that the variational DG approach does not determine the
wave function of an eigenstate (cf.~Sect.~V of GDF), 
no transport theory should attempt to do this; 
otherwise, the current would be identically zero unless the systems 
(e.~g., superconductors) sustain persistent currents.
We can find neither in Ref.~\onlinecite{DelaneyGreer:04a} nor elsewhere
a mention that the central variational DG ansatz would be an approximation. 
One deduces 
that the results are exact if one is able to consider sufficiently large sizes 
in \emph{full} CI calculations.
This is the case of the uncorrelated discrete model of I.
In a valid transport through uncorrelated systems 
genuinely based on the WF, the current conservation is exact
(provided that the Fourier completeness is satisfied if spatial and momentum grids 
are used).\cite{frensley:90} 
We demonstrated in I that the DG method 
does not automatically satisfy the current conservation, which needs be 
explicitly imposed. We did not misinterpret anything, as GDF argue,
we criticized the DG claim that this constraint is necessary only for other approaches, 
which use nonlocal interactions, truncate the molecular orbital basis set or the CI 
(second paragraph of Sect. VI in I), but \emph{not} for the DG method, as if that method 
were so good and automatically accounted for it. The current conservation is trivial, it is 
satisfied
by construction in the calculations where the current is constrained to be position independent,
as we also did in I. 
It is true ``\emph{that \ldots considerable care is needed in defining 
\underline{finite} expansions that are current conserving}'' (cf.~Sect.~V of GDF, underlined by us),
but this does not affect the results of I, which demonstrate that without explicit imposition,
the DG method violates the current conservation: they are deduced within the linear response limit
[$\mathcal{O}(V)$] and \emph{full} CI calculations. One should not confuse the two issues addressed in I: 
the violation of the current conservation, which is demonstrated by full CI calculations, and the 
very poor convergence, which is demonstrated by studying (as also done above) 
the effect of progressively increasing the number of 
\emph{exact eigen}states up to values beyond the reach of any feasible ab initio calculations.

In several places of their Comment, by
using the terms ``linear response approximation'' and 
``perturbation theory'', GDF indirectly mean criticism to I, e.~g.~in Sect.~V, 
where they mention ``\emph{the advantage of an analysis based upon an analytical model \ldots 
that \ldots avoids issues associated with linear response approximations, perturbation theory 
\ldots}''. GDF should have noted that, as emphasized in Sect.~\ref{sec-challenges}, 
their derivation of $g_0$ of Sect.~IV.B is also based on the linear response approximation:
the difference is that they employ it implicitly and heuristically, while our approach  
uses a systematic $\mathcal{O}(V)$ expansion spanning the whole Hilbert space (full CI).

The current oscillations mentioned by GDF in Sect.~V (which are rather \emph{irregular 
fluctuations} \cite{Baldea:unpublished}) have neither to do with the simplicity 
nor with the dimensionality ($d=1$) 
of the models of I. As noted there, for the uncorrelated model and the sizes 
considered by us, 
the time-dependent density matrix renormalization group (t-DMRG) 
yields the correct physical result, 
while the DG method lamentably fails either with or without imposing current conservation.
Likewise, the correct result is obtained within 
the standard Keldysh formalism,\cite{HaugJauho} despite the fact that 
the electrodes are modeled by the same tight-binding model used in I, and the employed 
Green functions $G(x, x^{\prime}, \varepsilon)$ 
should have been more affected by ``nearsightedness'' than the density matrix
$\rho(x, x^{\prime})$: they depend not only on 
the difference $x - x^{\prime}$, but also on energy. 
The exponent $d$ of the density matrix decay given by GDF is wrong; 
the correct one is $(1+d)/2$.\cite{Taraskin:02} 
GDF should have also noted that, 
for their $Au_{13}$- or $Au_{20}$-``electrodes'', each of $L/2\simeq 1$\,nm, 
the spatial variation of the density matrix within $\mathcal{L}\sim$$0.5$\,nm (Ref.~20) to which 
they refer is important, and the invoked ``nearsightedness'' problematic. Concerning this,
noteworthy, the above $\mathcal{L}$-value is deduced by calculating $\rho(x,x^{\prime})$
in very large systems of sizes much longer that $\mathcal{L}$, and not employing 
short DG-like electrodes, each of $L/2\simeq 1$\,nm.

The simple uncorrelated model of I, 
which is a textbook's example, correctly describes the major features of nanotransport,
if the well-established approaches mentioned above are applied,
but not within the DG method. 
Definitely, the failure is of the 
DG method itself and has nothing to do with the simplicity of the model. 
This issue regards the GDF model as well, 
for which their Landauer-type calculations and ours (cf.~Sect.~\ref{sec-challenges}) 
also yield the 
correct conductance, Kohn's principle notwithstanding. 
Of course, Kohn's principle is relevant for realistic systems, but,
curiously, GDF did not note that Kohn's tight-binding prediction provides the best overall 
fit of realistic calculations of the density matrix decay, as shown in the work to 
which they refer (Ref.~20).

Contrary to the GDF's claim, I is not a comment on the validity of the DG method.
In I we simply checked 
whether the DG method is able to describe the most simple 
uncorrelated and correlated systems.
If we wrote a comment, we would certainly have raised even 
more questions than on the fundamental issues 
of the next section. Letting alone minor issues (e.~g, the missing value of 
$V$ in Fig.~1 of Ref.~\onlinecite{DelaneyGreer:04a}), we would have asked, 
e.~g., how did DG conceive to apply their method
to the Kondo effect (cf.~last but one paragraph 
of Ref.~\onlinecite{DelaneyGreer:04a}). For typical 
Kondo temperatures $T_K\sim 1$\,mK$ - 1$\,K, this would imply 
to handle electrodes longer than the Kondo cloud length, 
$ L > \xi_K \sim \hbar v_F/(k_B T_K) \approx 10^3 - 10^6$\,nm within a CI expansion capable 
to very accurately describe a huge number of excited eigenstates, 
especially the very low excitations related to 
coherent spin fluctuations of energies $\sim k_B T_K \sim 10^{-4} - 10^{-7}$\,eV. 
\section{Discussion. Why does the variational DG method fail?}
\label{sec-fail}
As already stressed in Sect.~I, in I we criticized the WF-OBCs 
in the \emph{specific} context of the variational DG method (which means more than WF-OBCs) 
and \emph{not} otherwise.
Let us assume that GDF could have demonstrated that the WF-OBCs are correct.
What would be the implication on the (in)validity of the DG? \emph{None}, 
since our calculations were done \emph{just} by imposing WF-OBCs 
(i.~e., they were assumed to be ``correct''), and the fact that the results obtained 
within the DG approach 
presented in I are 
unphysical remains unaltered. The only implication would be that one could formulate more precisely:
the DG method fails not because the WF-OBCs \emph{per se} are wrong, but because,
with these OBCs, imposing one or more conditions prescribed by the DG to determine the steady 
state is unphysical. 

Even if (\emph{hypothetically}) examples could be found where results of a certain
approach (in our case, the DG's) were acceptable, a single counter-example suffices to demolish it.
The results for two examples presented in I as well as those for the GDF model of 
Sect.~\ref{sec-dg-to-gdf} unambiguously demonstrate the lamentable failure of the 
DG method both for uncorrelated and for uncorrelated transport. 
This is an irrefutable mathematical demonstration for the simplest 
correlated and uncorrelated, 
for discrete and continuum-space models, since their derivation uses 
\emph{nothing} else than the variational DG method prescribes. Consequently, the main objective 
of the present Reply has been achieved.

The problem that is really important 
is not whether the DG approach fails, but rather \emph{why} it fails. We do not present here 
a comprehensive analysis,\cite{Baldea:unpublished} but for the benefit of a reader interested 
in this method, we briefly note the following. Besides WF-OBCs, the DG  
method prescribes the total energy minimization 
and the usage of a normalized many-electron wave function $\Psi$ to describe 
a nonvanishing steady state current. 
Noteworthy, both conditions refer to a \emph{finite isolated} system.
So, within this philosophy, it could be possible to obtain a nonvanishing steady state 
current, i.~e., phenomenon that is manifestly irreversible, by merely examining a 
cluster which is not only finite
(and even very small, cf.~Ref.~\onlinecite{DelaneyGreer:04a}) but also 
\emph{isolated}, by imposing certain constraints at certain (very special) locations 
inside this cluster. 
The DG method uses \emph{absolutely no other information than that pertaining to a finite isolated system}:
this system is completely decoupled from the world, 
and there is absolutely no source of dissipation. 
At this point, it is important to emphasize that, basically, valid approaches to 
transport fall into two classes: 

 (i) Most widely used are transport theories, e.~g., 
based on the Keldysh NEGF or master equations, which also 
consider a finite cluster (wherein possible 
electron correlations are treated within DFT or more accurately \cite{Ng:88,Meir:92,HaugJauho}), but this finite cluster is linked to \emph{infinite} 
electrodes. It is this latter ingredient that accounts for irreversibility in a \emph{physically} 
justifiable manner: the 
imaginary parts of the embedding (contact) self-energies become nonvanishing \emph{only} in 
infinite electrodes.\cite{ElectrodeSelfEnergies} On the other side, 
that the DG approach 
can lead to a nonvanishing current is solely a lucky \emph{mathematical} consequence: 
certain matrix elements of the Fano operator (not directly related to an observable with an unequivocal 
physical meaning, like e.~g., the electronic number operator) 
computed somewhere inside of a finite isolated 
cluster happen to have nonvanishing imaginary parts, a fact \emph{by no means}  
related to a true physical dissipation.
Therefore, although \emph{by no means} critically 
related with the failure of the DG method demonstrated 
unambiguously mathematically, we reiterate our claim of I, that imposing WF-OBCs \emph{within}
the DG prescription is not physically justified. WF-OBCs can be used, but within \emph{other} approaches 
(e.~g., by solving the Liouville equation for the WF \cite{frensley:90}), for
which the present considerations do not in the least apply.
Consequence of an ad hoc mathematical constraint without a precise physical meaning,
the predicted DG current (vanishing or not) exhibits, not concidentally, 
completely unphysical trends (cf.~Sect.~\ref{sec-dg-to-gdf}, 
and Figs.~3 and 7 of I). Summarizing, in this paragraph we have indicated one
serious flaw of the DG method.

 (ii) Another class of transport treatments deduces the steady-state current by examining 
the long time ($t\to \infty$)
behavior of the wave function $\Psi(t)$ at zero temperature 
(e.~g., the already mentioned t-DMRG) or 
the statistical (density matrix) operator $\hat{\rho}(t)$ at finite temperatures.
\cite{AndreiLimits:06} In the former, starting from $x\sim -L/2$ (in the present notations) 
the many-body wave packet is monitored a sufficiently long time $t$, but 
before the packet reaches the other end ($x \sim +L/2$), since in the absence of any dissipation 
it will be reflected, a reversed current will appear, and current oscillations will last forever.
Mathematically, this amounts to derive steady-state properties (e.~g., electric current) by 
approaching the limit $L \to \infty$ \emph{first}, and only \emph{then} $t \to \infty$.
The clear physical analysis of Ref.~\onlinecite{AndreiLimits:06} explicitly emphasizes the 
importance of the correct order of these two limits for reaching the steady state; see Eq.~(6) 
there. This represents the counterpart of the fact well known in solid state physics on
the calculation of the dc-conductivity $\sigma_{dc}$ from the frequency- and 
wavevector-dependent conductivity $\sigma(\mathbf{q}, \omega)$. To obtain
the correct result, it is mandatory to take the limits $\mathbf{q} \to \mathbf{0}$ and 
$\omega \to 0$ (the counterparts of $L \to \infty$ and $t \to \infty$, respectively) 
in \emph{that} order, 
\[\sigma_{dc} = \lim_{\omega \to 0} \lim_{\mathbf{q} \to \mathbf{0}} 
\sigma(\mathbf{q}, \omega) ;
\]
see, e.~g., chapter 3.8 of Ref.~\onlinecite{mahan}.
The above considerations can be rephrased perhaps in a more direct 
manner as follows. The current operator $\hat{j}$ does not 
commute with the hermitian Hamiltonian $\hat{H}$ of a non-dissipative system, 
and consequently 
\[
\frac{\partial \hat{j}}{\partial t} = i [\hat{H}(t), \hat{j}(t)] \neq 0 .
\] 
A steady state $\lim_{t\to \infty}\Psi(t)$ characterized by a 
stationary current 
\[\lim_{t\to \infty} \partial I(t)/\partial t = 0 , \mbox{ and } 
\lim_{t\to \infty} I(t) = I \neq 0
\]
can only be obtained because, after averaging the commutator
(a nonvanishing operator),
$\langle \Psi \vert [\hat{H}, \hat{j}]\vert \Psi\rangle_{t}$, 
the average can vanish if one first takes the infinite 
``volume'' limit $L\to \infty$ and then $t \to \infty$. 

In examining a steady state transport, two general principles of the 
thermodynamics of irreversible processes are mostly discussed in the literature: 
the minimum entropy production \cite{prigogine:49} and the 
entropy maximization.\cite{Jaynes:57b} Prigogine's discussion of the steady
state within the former principle \cite{prigogine:49} clearly revealed that the 
aforementioned order of the two limits is essential. Within the same principle, 
the same fact was nicely illustrated in the particular case of the flow 
through a capillary tube connecting two containers with ideal gas at different 
pressures.\cite{MinEntropyProduction}
Were the system finite, the flow would be no more irreversible: after some time, the gas 
would flow back to higher pressure. 
In Ref.~\onlinecite{DelaneyGreer:04a}, 
DG claimed that they can deduce the variational ansatz,
which they used to compute the steady state wave function $\Psi$,
from the maximum entropy principle. Except for citing Ref.~\onlinecite{Jaynes:57a}, 
neither in Ref.~\onlinecite{DelaneyGreer:04a}, nor in later 
works,\cite{DelaneyGreer:04b,DelaneyGreer:06,Fagas:07} or in the Comment 
any detail on this derivation was provided. 
In Ref.~\onlinecite{Jaynes:57a}, Jaynes 
presented quantitative considerations emerging from that principle 
applied to systems of very large number of degrees of freedom based on Shannon's entropy 
for statistical equilibrium;
a steady state is not even mentioned. In a subsequent paper,\cite{Jaynes:57b}
cited in a later work by DG,\cite{DelaneyGreer:06} Jaynes further analyzed 
the time-dependent case in detail from the perspective of 
a single basic principle (entropy maximization) applied to all cases, 
equilibrium or otherwise. Again, the case of a steady state was not explicitly 
considered,
and consequently the issue of the correct limit order to reach a steady state 
within the principle of entropy maximization was not addressed. 
Prior to the DG work, there has 
been attempted to recover the standard Landauer results from the maximum entropy 
inference,\cite{Bokes:03} and it turned out that schemes based on that principle 
encounter notable difficulties \emph{even} if the limits $L \to \infty$ and $t \to \infty$ 
are taken in the correct order. Ignoring these difficulties and without validating their 
variational scheme against any well-established theoretical result (as done by us in I), 
DG put forward an approach, wherein, implicitly, the limit order is exactly \emph{opposite} to  
the correct one: they consider a time-independent wave function $\Psi$
(amounting to take $t\to \infty$), and then (by taking the largest cluster it can handle, 
which is in fact very small) attempt to mimic the limit $L \to \infty$.
This was the \emph{physical} reason why in I we employed the acronym SWF (stationary Wigner function)
for what we now called the DG method. Therefore, within this physical context we do not think,
contrary to GDF (cf.~their Ref.~4), that the term SWF is misleading. Summarizing, in this paragraph
we have indicated another serious flaw of the DG method, 
which represents the \emph{fundamental physical reason} why it fails.
\section{Conclusion}
\label{sec-conclusion}
GDF seem to have realized that their claim is wrong; 
in Sect.~III, they write that our conclusion 
``\emph{that an asymmetric injection of electrons is needed to obtain 
a current is incorrect, \underline{if} injection refers to incoming electron \emph{momentum} 
distributions\ldots}'' (underlined by us).
However, the above conditional clause does not apply and therefore 
our critique of I is not in the least affected. 

To conclude, in this Reply we have 
demonstrated that in their Comment GDF could neither show that the critique is incorrect, 
nor give even a single example where the DG method can correctly describe a transport property. 
We used their model 
to complete the evidence on the failure of the DG method presented in I. 
Letting alone the fundamental reasons against this method (cf.~Sect.~\ref{sec-fail}), 
because the failure of the DG method, 
\cite{DelaneyGreer:04a,DelaneyGreer:04b,DelaneyGreer:06}
which is a \emph{unequivocal mathematical} prescription, 
comprises the simplest uncorrelated and correlated systems described 
within discrete (Ref.~\onlinecite{Baldea:2008b}) and continuum (Sect.~\ref{sec-dg-to-gdf}) 
spaces, it would be a utopia to presume that real systems could be correctly described. 

Most importantly for readers interesting in
the DG method, we have presented not only further exact  
results showing 
that this method fails, but also indicated the basic physical reasons why it fails.

DG argued that their essential ingredient, the variational ansatz, 
is deduced from the 
maximum entropy principle. In Ref.~\onlinecite{Jaynes:57b} cited by DG,\cite{DelaneyGreer:06} the 
author asserted that
``\emph{if it can be shown that the class of phenomena predictable by maximum-entropy 
inference differs in any way from the class of experimentally reproducible phenomena, 
that fact would demonstrate the existence of new laws of physics, not presently known.}''
This statement may apply to the results deduced within the DG method: neither 
the prediction of the present Sect.~\ref{sec-dg-to-gdf} (a vanishing conductance on resonance),
nor a conductance increasing if one moves away from resonance (Fig.~3 of I), or conductance
maxima of the Coulomb blockade peaks becoming higher and even exceeding $g_0$ 
with decreasing dot-electrode coupling 
$t_d$ (Fig.~7 of I) have been observed so far. These trends are just opposite 
to those of the available experiments\ldots, so should one still await the advent of new 
physical laws in the sense quoted above from Ref.~\onlinecite{Jaynes:57b}? 
Until then, one should not be too surprised if the DG predictions are also at odds with
other existing experiments of molecular electronics. Until recently,\cite{Reed:09} 
although being unable to demonstrate that their theory is sound,
DG could claim that their method \cite{DelaneyGreer:04a} produces current values better agreeing 
with experiment than the other theoretical estimations.
The accurate data of the beautiful experiment of Ref.~\onlinecite{Reed:09}
have clearly demonstrated that just the \emph{opposite} is true. To see this, one can simply
compare Fig.~2a of Ref.~\onlinecite{Reed:09}, Fig.~2 of Ref.~\onlinecite{DelaneyGreer:04a},
and, e.~g., Fig.~4 of Ref.~\onlinecite{Tomfohr:04} among themselves.
For instance, at $V=1; 1.5$\,volt, the currents in $\mu$A are $\simeq 1.7; 6$,\cite{Reed:09}
$\simeq 0.09; 0.13$,\cite{DelaneyGreer:04a} and $\simeq 6; 9.7$.\cite{Tomfohr:04}
So, without any special fine tuning (e.~g., contact geometry), 
a standard NEGF-DFT approach \cite{Tomfohr:04} 
can reasonably describe the experimental data, while the DG's cannot. 

As a matter of principle, we end by reiterating that 
whether (un)luckily results obtained within a certain theoretical 
(DG's or whatsoever) method (dis)agree with experiment is \emph{neither} the only 
\emph{nor} the decisive 
point to assess its validity: a comparison with experiment is meaningful 
only if its physical basis is sound. 
The authors acknowledge the financial support for this work
provided by the Deu\-tsche For\-schungs\-ge\-mein\-schaft.
\bibliographystyle{aip}

\begin{thebibliography}{10}
\bibitem{round-L}
The employed numerical $L$-values do not represent special choices; their
  slight deviations from an integer only reflect the fact that $k_F L /\pi$
  (electron number) should be an integer.
\bibitem{DG-numerical}
Unlike in the sound Landauer approach, wherein the linear and nonlinear
  response can be deduced analytically, DG-calculations must be performed
  numerically even within the linear response limit.
\bibitem{x-p-grids}
Attempts to change the $p$- and $x$-grids used to constrain the boundary WF 
  and the current only yield irrelevant changes in the
  DG-conductance. For various grid choices [uniform grids, omission of some
  $p$-points around $p=0$ or where $f_{0}(q_{L,R}, p)$ ``wildly'' varies,
  use of the grids indicated by Frensley \cite{frensley:90} to satisfy the
  Fourier completeness (although not necessary, because
  integrations over the continuous $x$ are done
  analytically), etc], one gets $g_{DG}$-values ranging from $\sim 10^{-13} -
  10^{-7} g_0$.
\end{thebibliography}
\end{document}